\documentclass{camera}
\usepackage{amsmath}
\usepackage{epsfig}

\def\as{\alpha_{\mbox{\scriptsize s}}}

\def\eps{\epsilon}

\def\cF{{\cal{F}}}

\def\Journal#1#2#3#4{{#1} {\bf #2} (#3) #4}

\def\EPJC{{\em Eur. Phys. J.} C}

\def\JHEP{{\em J. High Energy Phys.}}
\def\JPG{{\em J. Phys.} G}

\def\NPB{{\em Nucl. Phys.} B}
\def\PLB{{\em Phys. Lett.}  B}

\begin{document}

%
\title{SEMI-NUMERICAL RESUMMATION OF EVENT SHAPES AND JET RATES}

%
\author{Andrea Banfi}

%
\organization{Universit\`a di Milano-Bicocca and INFN, Sezione di
  Milano, Italy}

\maketitle

\abstract{We describe a numerical procedure to resum multiple emission
  effects in event shape variables and jet rates.}

%
\section{Introduction}
\label{sec:intro}
Event shape variables (and jet rates) are among the most studied
observables in QCD.  They are useful both to measure $\as$
\cite{Bethke} and to search for genuine non perturbative (NP) effects
\cite{NPstandards}. A perturbative (PT) fixed order expansion
describes well an event shape rate (the fraction $\Sigma(v)$ of events
whose value of the event shape $V$ is less than $v$) in the region
$v\sim 1$. In the less inclusive region $v\ll 1$ large logarithms
$\as^m \ln^n v$ arising from incomplete real-virtual cancellations
have to be resummed to all orders to give meaning to the PT series
[3-7].
In order to fix the scale of $\as$ one needs to control in $\ln
\Sigma$ all terms of the form $\as^n L^{n+1}$ (leading LL or double DL
logarithms) and $\as^n L^n$ (next-to-leading NLL or single SL
logarithms), with $L=\ln 1/v$.  It is therefore important to understand
the sources of large logarithmic contributions and to find a general
way to resum them, at least at NLL level. The aim of the paper is to
describe a general procedure to resum SL terms arising from multiple
emission effects, which is the most difficult task in any resummation
programme.

\section{Classification of large logarithms}
\label{sec:res}
Leading logarithms originate from soft and collinear gluon radiation
and are in general straightforward to resum. SL contributions
instead have a variety of sources:
\begin{itemize}
\item running of the coupling;
\item soft emission at large angles;
\item hard collinear splitting;
\item multiple emission effects.
\end{itemize}
The first three contributions can be easily resummed by computing the
first order result (in the soft-collinear limit), and exponentiating
the answer.  The last contribution is the most difficult to address.
In the luckiest case (additive observable) it requires only one Laplace
transform \cite{CTTW}, in more complicated cases one needs to
introduce an additional amount of Fourier transforms \cite{NewBroad}
(which can be as many as five in the thrust minor case \cite{Tmin}).
There are situations (thrust major, oblateness) in which an analytical
formulation does not even exist at all. It is therefore important to
investigate if there is a general method capable to resum SL terms
arising from multiple emission effects.

\section{Resummation of multiple emission effects}
\label{sec:multi}
A resummed integrated distribution for a `suitable' (for a definition
see \cite{numsum}) observable $V$ can be written in the form
\begin{equation}
  \label{eq:res-dist}
  \Sigma(v)=e^{-R(v)}\>\cF(R')\>,\qquad R'=-v\frac{dR}{dv}\>,
\end{equation}
where the radiator $R$ builds up the Sudakov form factor obtained by
exponentiating the one soft-collinear gluon contribution, while
$\cF(R')$ is a SL function which accounts for multiple emission effects.
Suppose that we know the resummed distribution $D_s(v_s)$ of a
`simple' variable $V_s$. The distribution $D(v)$ of a more complicated
observable $V$ is related to $D_s(v_s)$ by:
\begin{equation}
  \label{eq:simple}
  D(v)=v\int\frac{dv_s}{v_s}P(v|v_s)\>D_s(v_s)\>,
\end{equation}
with $P(v|v_s)$ the probability of having $V$ equal to $v$ given a
value $v_s$ for $V_s$. The result \eqref{eq:simple} is quite general.
If $V$ and $V_s$ have the same DL structure, it can be simplified to
give, at SL accuracy:
\begin{equation}
  \label{eq:F}
  \begin{split}
    \Sigma(v)=\Sigma_s(v)\,\cF(R')\>,\quad
    \cF(R')=\int\frac{dx}{x}x^{-R'}p(x,R')\>,\quad 
    p\left(\frac{v}{v_s},R'\right)=v P(v|v_s)\>.
  \end{split}
\end{equation}
If we now choose
$V_s(\{k_i\})=\max_i\{V(k_i)\}$, the resummation of
$\Sigma_s(v_s)$ leads directly to
\begin{equation}
  \label{eq:res-simple}
  \Sigma_s(v_s)=e^{-R(v_s)}\>,
\end{equation}
so that the function $\cF(R')$ obtained from \eqref{eq:F} coincides
with the one introduced in \eqref{eq:res-dist}.

The effort of any resummation is then to compute $p(x,R')$. This can
be done numerically with the following  Monte Carlo procedure.
One starts with a given Born configuration with a fixed number of
jets. One then
\begin{enumerate}
\item[0] fixes a value $v_s$ of the simple observable;
\item[I] generates a soft and collinear emission according to the
  phase space $R'$, with $V(k_i)<V(k_{i-1})$;
\item[II] if $V(k_i)<\eps \ll 1$ one stops, otherwise goes back to
  step $I$;
\item[III] given the momentum set $\{k_i\}_{i=1,\dots,n}$ one computes the
  value of the observable $v=V(k_1,\dots,k_n)$.
\end{enumerate}
The above procedure gives the probability $P(v|v_s)$, which can be
integrated according to \eqref{eq:F} to give the function $\cF(R')$.  

\begin{figure}[htp]
  \begin{center}
    \epsfig{file=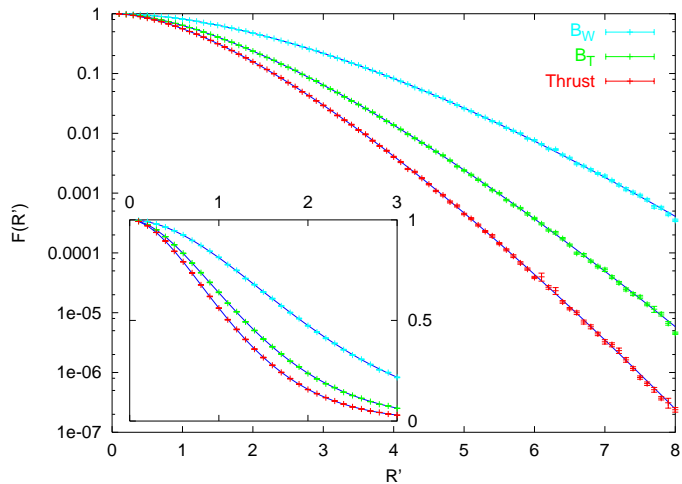, width=0.6\textwidth}
    \caption{Comparison of the analytical and numerical results for the
      function $\cF(R')$ for the thrust and the two broadenings. The
      lines represent the analytical predictions while the dots are
      the numerical results.}
    \label{fig:old}
  \end{center}
\end{figure}
\begin{figure}[htp]
  \begin{center}
    \epsfig{file=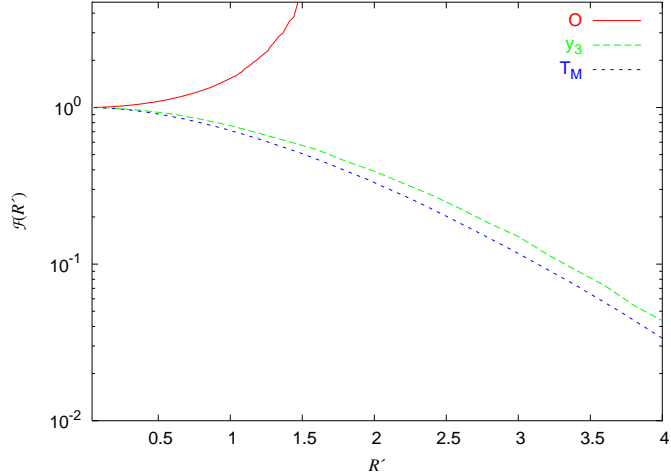, width=0.6\textwidth}
    \caption{The function $\cF(R')$ for the thrust major $T_M$, the
      oblateness $O$ and the three-jet resolution $y_3$. }
    \label{fig:new}
  \end{center}
\end{figure}

The method is process independent and has been tested in the case of
observables for which an analytical resummation is known. For the thrust
and the two broadenings $B_T$ and $B_W$ the results are shown in
figure \ref{fig:old}. The Monte Carlo procedure reproduces with great
precision the function $\cF(R')$ for these observables. This means
that numerical results are truly interchangeable with analytical
ones. The method has then been applied \cite{numsum} to the thrust
major $T_M$ and the oblateness $O$, for which a resummed prediction
did not exist yet, and to the three-jet resolution $y_3$ (Durham
algorithm \cite{JetRates}), for which only a part of SL contributions
had been computed \cite{DissSchmell} (see figure \ref{fig:new}).
This new method can then be seen as a first step towards a full automatic
resummation of event shapes and jet rates at NLL accuracy.

\paragraph{Acknowledgements}
It was a pleasure to carry out this work with Gavin Salam and Giulia
Zanderighi. I am also grateful to Matteo Cacciari, Yuri Dokshitzer,
Pino Marchesini, Lorenzo Magnea and Graham Smye for helpful comments and
suggestions.

%
\end{document}